\documentclass[conference]{IEEEtran}
\IEEEoverridecommandlockouts
\usepackage{cite}
\usepackage{amsmath,amssymb,amsfonts}
\usepackage{algorithmic}
\usepackage{graphicx}
\usepackage{textcomp}
\usepackage{xcolor}
\def\BibTeX{{\rm B\kern-.05em{\sc i\kern-.025em b}\kern-.08em
    T\kern-.1667em\lower.7ex\hbox{E}\kern-.125emX}}
\begin{document}

\title{A Data-Driven Prescribed-Time Control Framework via Koopman Operator and Adaptive Backstepping*\\
{\footnotesize \textsuperscript{*}Note: Sub-titles are not captured in Xplore and
should not be used}
\thanks{This work was supported by... (Please insert your funding information here).}
}

\author{\IEEEauthorblockN{Yue Wu}
\IEEEauthorblockA{\textit{School of Automation Science and Engineering, Xi'an Jiaotong University}\\
Xi'an, 710049, China \\
\textit{Xinjiang Cigarette Factory, Hongyun Honghe Tobacco (Group) Co., Ltd.}\\
Urumqi, 830000, China \\
Email: wuyue0619@stu.xjtu.edu.cn}
}

\maketitle

\begin{abstract}
Achieving rapid and time-deterministic stabilization for complex systems characterized by strong nonlinearities and parametric uncertainties presents a significant challenge. Traditional model-based control relies on precise system models, whereas purely data-driven methods often lack formal stability guarantees, limiting their applicability in safety-critical systems. This paper proposes a novel control framework that synergistically integrates data-driven modeling with model-based control. The framework first employs the Extended Dynamic Mode Decomposition with Control (EDMDc) to identify a high-dimensional Koopman linear model and quantify its bounded uncertainty from data. Subsequently, a novel Prescribed-Time Adaptive Backstepping (PTAB) controller is synthesized based on this data-driven model. The design leverages the structural advantages of Koopman linearization to systematically handle model errors and circumvent the "explosion of complexity" issue inherent in traditional backstepping. The proposed controller is validated through simulations on the classic Van der Pol oscillator. The results demonstrate that the controller can precisely stabilize the system states to a small neighborhood of the origin within a user-prescribed time, regardless of the initial conditions, while ensuring the boundedness of all closed-loop signals. This research successfully combines the flexibility of data-driven approaches with the rigor of Lyapunov-based analysis. It provides a high-performance control strategy with quantifiable performance and pre-assignable settling time for nonlinear systems, showcasing its great potential for controlling complex dynamics.
\end{abstract}


\section{Introduction}
\label{sec:introduction}

\subsection{Motivation and Challenges}
Modern engineering systems, such as robotics, aerospace, and power systems, are characterized by increasing complexity, inherent nonlinearities, and significant operational uncertainties \cite{hua2022}. The precise control of these systems is paramount to achieving high performance, safety, and reliability. Within control theory, two dominant paradigms have emerged to address these challenges: model-based control and data-driven control \cite{li2025data}.

Model-based approaches, including adaptive and robust control, rely on an accurate mathematical description of the system dynamics. When an accurate model is available, these methods provide strong theoretical guarantees and superior performance. However, for many complex systems, deriving a model that is both accurate and tractable for controller design via first-principles is a formidable, if not impossible, task \cite{yan2025}.

Conversely, the rise of data-driven control offers a path to bypass intricate modeling procedures \cite{castellanos2024}. These methods learn controllers or system behaviors directly from input-output data, showing immense potential for systems with unknown or highly uncertain models \cite{zhang2025}. Nevertheless, purely data-driven methods, especially those considered "black-boxes," often lack rigorous stability and performance guarantees, rendering them unacceptable for safety-critical applications \cite{huang2024}.

This dichotomy between model-based and data-driven approaches raises a pivotal question: can we synergistically combine their strengths? The literature presents an ongoing debate. Some argue that data-driven methods can outperform model-based counterparts by avoiding undermodeling issues \cite{prag2022}. Others view data-driven techniques as tools to enhance or construct models within a model-based framework \cite{alhajeri2022}. This divergence points to a deeper need in control engineering: a unified framework that fuses data with physical models. A purely black-box data-driven controller is untenable in safety-critical systems due to its lack of interpretability and formal guarantees \cite{huang2024}. Conversely, a controller based on an imperfect model suffers from robustness issues due to unavoidable modeling errors \cite{aitabbas2023}.

Therefore, the most promising path forward is to leverage data to construct a model that possesses a specific structure amenable to formal analysis. The core idea of this paper is precisely this: instead of building a black-box controller, we utilize Koopman operator theory to learn a linear model with quantifiable uncertainty bounds from data \cite{bold2025}. This learned model then serves as the basis for a rigorous, Lyapunov-based controller design \cite{narasingam2019}. This integrated strategy bridges the gap between the flexibility of data-driven methods and the rigor of model-based control, offering a powerful framework to tackle modern control challenges.

\subsection{State of the Art and Its Limitations}
Traditional nonlinear control techniques, such as adaptive backstepping, have achieved great success in handling systems with structured uncertainties \cite{cong2023}. However, these methods critically depend on the system conforming to a specific structural form, like the strict-feedback form \cite{singh2023}, and their stability guarantees are typically limited to asymptotic stability \cite{ao2021}, where states converge to the desired value only as time approaches infinity.

In pursuit of superior control performance, researchers have proposed more advanced control objectives. Finite-time and fixed-time stability theories improve upon the convergence rate of asymptotic stability \cite{ai2023}, but their settling times often depend on initial conditions or non-adjustable design parameters, a significant limitation in tasks requiring precise temporal coordination.

To overcome this limitation, **Prescribed-Time Control (PTC)** has emerged as a state-of-the-art benchmark for control performance \cite{badiei2024}. PTC guarantees that the system's settling time, $T$, is a user-defined parameter that is completely independent of initial conditions and any other design parameters \cite{ding2024}.

Simultaneously, in the realm of data-driven modeling, Koopman operator theory has garnered widespread attention as a powerful tool for global linearization of nonlinear systems \cite{bevanda2021}. This theory enables a nonlinear dynamical system to be described by a linear Koopman operator by lifting the dynamics to an infinite-dimensional function space. This property makes it an ideal candidate for data-driven modeling.

Despite significant advances in both PTC and Koopman theory, combining them to design a control framework that can both learn a model from data and guarantee prescribed-time convergence remains an open and challenging research problem.

\subsection{Paper's Thesis and Contributions}
This paper develops a novel data-driven control framework for a class of strict-feedback nonlinear systems that achieves prescribed-time tracking. The framework synergistically integrates: (1) the Extended Dynamic Mode Decomposition with Control (EDMDc) method for learning a linear representation of the system dynamics and quantifying its error bounds \cite{schaller2023}, and (2) a novel Prescribed-Time Adaptive Backstepping (PTAB) controller synthesized based on the learned model \cite{cong2023}. The stability and performance of this controller are formally proven through a comprehensive Lyapunov analysis.

\textbf{Major Contributions}:
\begin{itemize}
    \item \textbf{Novel Controller Synthesis}: A PTAB controller is designed that explicitly incorporates the structure of the data-driven Koopman model and its bounded uncertainty.
    \item \textbf{Synergistic Problem Solving}: We demonstrate how Koopman linearization naturally mitigates the "explosion of complexity" problem inherent in traditional backstepping.
    \item \textbf{Rigorous Stability Proof}: A complete Lyapunov-based proof is provided, which formally accounts for the approximation error of the data-driven model, guaranteeing prescribed-time stability and the boundedness of all closed-loop signals.
\end{itemize}

To clearly position our contributions, Table \ref{tab:comparison} provides a qualitative comparison of relevant control methodologies.

\begin{table*}[t] 
\centering
\caption{Qualitative Comparison of Control Methodologies}
\label{tab:comparison}
\begin{tabular}{|c|c|c|c|}
\hline
\textbf{Feature} & \textbf{Traditional AB} & \textbf{Standard Koopman MPC} & \textbf{Proposed Koopman-PTAB} \\
\hline
\textbf{Conv. Type} & Asymptotic & Typically Asymptotic & \textbf{Prescribed-Time} \\
\hline
\textbf{Conv. Time} & IC/Gain Dependent & Tuning Dependent & \textbf{User-Prescribed} \\
\hline
\textbf{Nonlinearity} & Recursive Design & Global Linearization & \textbf{Integrated Design} \\
\hline
\textbf{Data Need} & None (Model-Based) & Required & \textbf{Required (with bounds)} \\
\hline
\textbf{Guarantees} & Lyapunov Stability & Recursive Feasibility & \textbf{PTS Lyapunov Stability} \\
\hline
\end{tabular}
\end{table*}

\section{Preliminaries and Problem Formulation}

\subsection{System Class: Parametric Strict-Feedback Systems}
We consider a class of nonlinear systems whose dynamics can be described in the parametric strict-feedback form, which is a standard structure for backstepping design:
\begin{equation} \label{eq:sys_model}
\begin{cases}
    \dot{x}_1 &= x_2 + f_1(x_1) + \theta^T \phi_1(x_1) \\
    \dot{x}_2 &= x_3 + f_2(x_1, x_2) + \theta^T \phi_2(x_1, x_2) \\
    & \vdots \\
    \dot{x}_n &= u + f_n(x) + \theta^T \phi_n(x) + d(t)
\end{cases}
\end{equation}
where $x = [x_1, \dots, x_n]^T \in \mathbb{R}^n$ is the state vector, $u \in \mathbb{R}$ is the control input, $\theta \in \mathbb{R}^p$ is a vector of unknown constant parameters, $f_i(\cdot)$ and $\phi_i(\cdot)$ are known nonlinear functions, and $d(t)$ is a bounded external disturbance satisfying $|d(t)| \leq d_{\max}$ for a known positive constant $d_{\max}$.

\subsection{Data-Driven Modeling via Koopman Operator Theory}
Koopman operator theory provides a linear perspective for analyzing nonlinear dynamics. For a continuous-time nonlinear system described by $\dot{x} = F(x)$, the Koopman operator $\mathcal{K}$ is an infinite-dimensional linear operator acting on a space of observables $g(x)$. The evolution of these observables is governed by:
\begin{equation}\label{eq:koopman_def}
    \frac{d}{dt}g(x(t)) = (\mathcal{K}g)(x(t))
\end{equation}
In practice, a finite-dimensional approximation is necessary. We introduce a dictionary of $N$ basis functions, $\Psi(x) = [\psi_1(x), \dots, \psi_N(x)]^T$. Using EDMDc from a set of $M+1$ data snapshots $\{x_k, u_k\}_{k=0}^M$, we seek the best-fit linear model in the lifted space by solving the least-squares problem:
\begin{equation}\label{eq:edmdc_ls}
    \min_{A_d, B_d} \sum_{k=0}^{M-1} ||\Psi(x_{k+1}) - A_d \Psi(x_k) - B_d u_k||_2^2
\end{equation}
The solution yields a discrete-time model, which can be converted to its continuous-time counterpart:
\begin{equation}\label{eq:koopman_continuous}
    \dot{z} = Az + Bu + r(t)
\end{equation}
where $z = \Psi(x)$, and $r(t)$ represents the residual or approximation error. The original state can be approximately reconstructed via a matrix $C$, i.e., $x \approx Cz$. The accuracy of this model is subject to projection errors from the choice of dictionary and estimation errors from finite, noisy data \cite{kanai2022}.

Our core modeling philosophy is not to assume a perfect linear model but to explicitly model the lifted system as a linear system with \textbf{bounded uncertainty}. We lump all sources of uncertainty—the disturbance $d(t)$, the Koopman approximation error $r(t)$, and the state reconstruction error—into a single term $\Delta$. The problem is thus refined to:
\begin{equation}\label{eq:lifted_uncertain}
    \dot{z} = Az + Bu + \Delta(z, u, t)
\end{equation}
We make the following crucial assumption.

\textbf{Assumption 1 (Uncertainty Bound):} The total uncertainty $\Delta$ is bounded by a known linear function of the lifted state norm:
\begin{equation}\label{eq:uncertainty_bound}
    ||\Delta(z, u, t)|| \leq \delta_0 + \delta_1 ||z||
\end{equation}
where $\delta_0, \delta_1 > 0$ are constants that can be estimated from a validation dataset. This is a standard and reasonable assumption in robust adaptive control \cite{kanai2022}, as the states of a controlled system typically operate within a compact set. These parameters can be estimated by computing the one-step-ahead prediction error $\Delta_k = \frac{\Psi(x_{k+1}) - \Psi(x_k)}{\Delta t} - (A\Psi(x_k) + Bu_k)$ on a validation set and finding a conservative linear bound over the data pairs $(||\Delta_k||, ||\Psi(x_k)||)$. This formalization of uncertainty is key to bridging the data-driven modeling step with the subsequent robust controller design \cite{lu2023}.

\subsection{Prescribed-Time Stability (PTS)}
\textbf{Definition 1 (PTS):} An equilibrium point is said to be prescribed-time stable if it is fixed-time stable and its settling time $T$ can be arbitrarily pre-assigned by the user during the design stage \cite{li2025data}.

The core mechanism for achieving PTS is a time-varying scaling function that grows to infinity as $t \to T$. We employ the standard gain function:
\begin{equation}\label{eq:pts_gain}
    \mu(t) = \frac{T^2}{(T-t)^2}, \quad t \in [0, T)
\end{equation}

\textbf{Definition 2 (PPTS):} Given the presence of uncertainties, exact convergence to the origin may be infeasible. Therefore, our objective is Practically Prescribed-Time Stability (PPTS) \cite{shakouri2022}, which guarantees that the system trajectories enter and remain within a small residual set around the origin for all $t \geq T$.

\subsection{Control Objective}
The control objective is formally stated as: for the nonlinear system \eqref{eq:sys_model}, a desired reference trajectory $x_d(t)$, and a user-specified time $T > 0$, design an adaptive control law $u(t)$ and parameter update law $\dot{\hat{\theta}}$ such that for any initial condition $x(0)$, the tracking error $e(t) = x(t) - x_d(t)$ converges to a small, computable residual set $\Omega_e$ for all $t \geq T$. All signals in the closed-loop system must remain bounded.

\section{Data-Driven PTAB Controller Design}
\subsection{Design Philosophy}
The core idea is to apply the recursive backstepping methodology to the high-dimensional, data-driven \textit{linear} representation of the system, not the original nonlinear dynamics. This fundamentally simplifies the design and circumvents two major obstacles of traditional backstepping. First, it obviates the need for the learned matrices $A$ and $B$ to have a specific sparse structure, as the entire $Az$ term is treated as a feedforward component. Second, it naturally mitigates the "explosion of complexity" \cite{umutlu2024}, as the derivatives of virtual control laws are structurally simple due to the linearity of the learned model.

\begin{figure}[htbp]
\centering
\includegraphics[width=0.8\columnwidth]{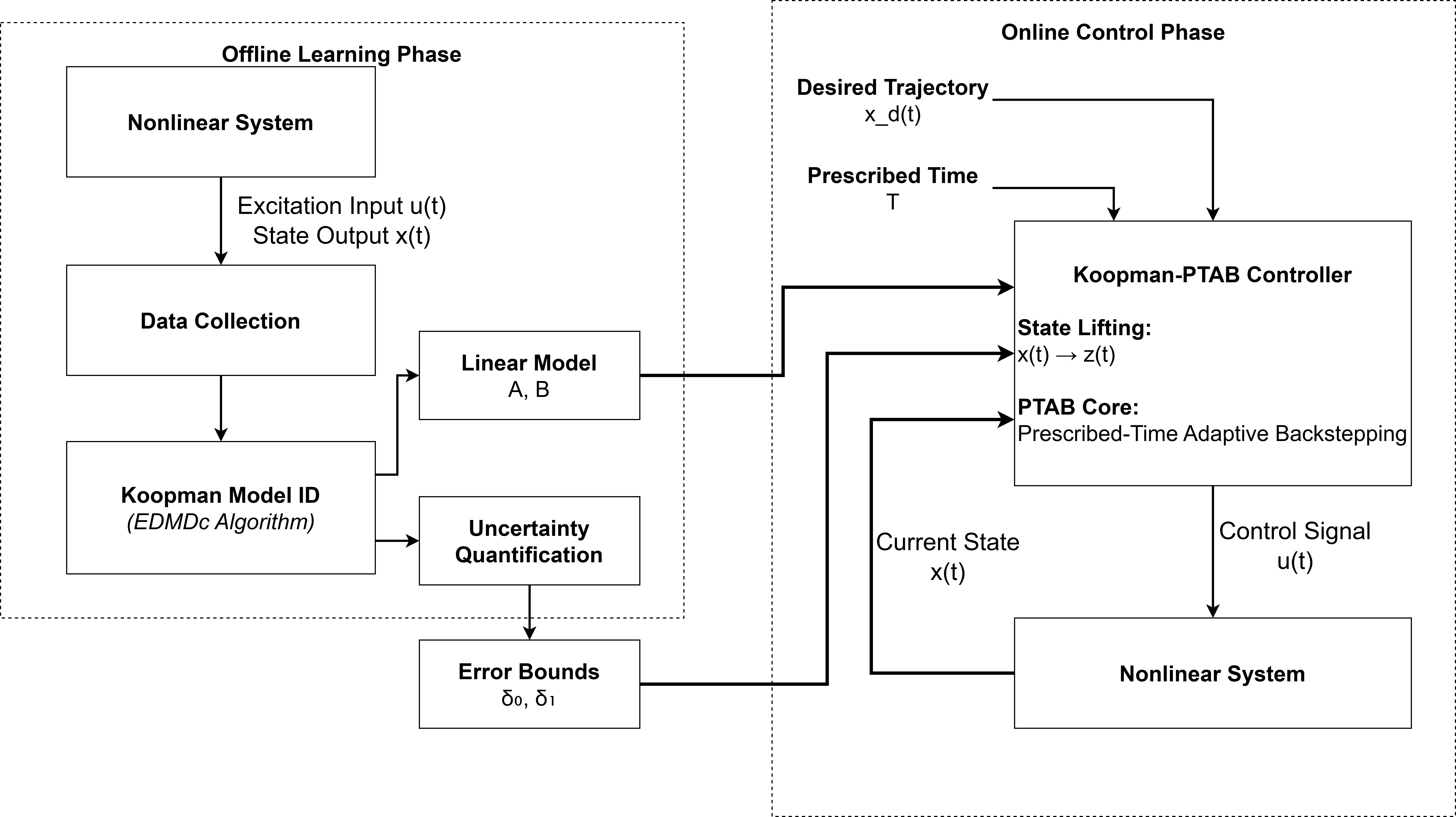} 
\caption{Block diagram of the proposed Koopman-PTAB control architecture, highlighting the synergy between data-driven modeling and backstepping design.}
\label{fig:control_diag}
\end{figure}
The overall architecture of the proposed data-driven prescribed-time control framework is illustrated in Fig.~\ref{fig:control_diag}. The framework is distinctly separated into two phases: an offline learning phase and an online control phase.

In the Offline Learning Phase, the primary goal is to derive a tractable linear model with quantifiable uncertainty from the complex nonlinear system. This is achieved by first applying a rich excitation input $u(t)$ to the system to generate a comprehensive dataset of state trajectories $x(t)$. The collected data is then processed by the Extended Dynamic Mode Decomposition with Control (EDMDc) algorithm to identify a high-dimensional linear model, represented by matrices $A$ and $B$. Crucially, a parallel process of Uncertainty Quantification is performed. By comparing the predictions of the learned linear model against a separate validation dataset, we establish conservative error bounds, characterized by the parameters $\delta_0$ and $\delta_1$. These bounds encapsulate all modeling inaccuracies and external disturbances.

In the Online Control Phase, the pre-computed linear model and its error bounds serve as the foundation for the real-time controller. The Koopman-PTAB Controller receives the desired trajectory $x_d(t)$, the user-defined prescribed time $T$, and the current state of the system $x(t)$ as inputs. Internally, the controller first performs a State Lifting operation, mapping the current state $x(t)$ into the high-dimensional Koopman space, $z(t)$. The core of the controller, the PTAB Core, then utilizes the principles of Prescribed-Time Adaptive Backstepping, leveraging the linear structure of the learned model and the quantified error bounds, to compute the necessary control signal $u(t)$. This signal is then applied to the actual nonlinear system, closing the feedback loop and driving the system to follow the desired trajectory within the prescribed time.

\subsection{Recursive Controller Derivation}
We define error coordinates in the $N$-dimensional lifted space. Let $z = \Psi(x)$ and $z_d = \Psi(x_d)$. The error variables are:
\begin{equation}\label{eq:error_coords}
\begin{aligned}
    e_1 &= z_1 - z_{d,1} \\
    e_i &= z_i - \alpha_{i-1}, \quad i = 2, \dots, N
\end{aligned}
\end{equation}
where $\alpha_{i-1}$ is the virtual control law for the $(i-1)$-th subsystem. For simplicity, let $\rho(t) = 2/(T-t)$.

\textbf{Step 1:} Consider the Lyapunov function $V_1 = \frac{1}{2} e_1^2$. Its derivative is $\dot{V}_1 = e_1 \dot{e}_1 = e_1((Az)_1 + (Bu)_1 + \Delta_1 - \dot{z}_{d,1})$. We design the virtual control law $\alpha_1$ for $z_2$ to stabilize this subsystem:
\begin{equation}\label{eq:alpha1}
    \alpha_1 = -c_1 \rho(t) e_1 - (Az)_1 - (Bu)_1 + \dot{z}_{d,1}
\end{equation}
where $c_1 > 0$. Substituting $z_2 = e_2 + \alpha_1$ into $\dot{V}_1$ yields:
\begin{equation}\label{eq:V1_dot}
    \dot{V}_1 = -c_1 \rho(t) e_1^2 + e_1 e_2 + e_1 \Delta_1
\end{equation}

\textbf{Step i (Recursive Step, $2 \le i < N$):} Consider the Lyapunov function $V_i = V_{i-1} + \frac{1}{2} e_i^2$. The virtual control law $\alpha_i$ is designed as:
\begin{equation}\label{eq:alphai}
    \alpha_i = -c_i \rho(t) e_i - e_{i-1} - (Az)_i - (Bu)_i + \dot{\alpha}_{i-1}
\end{equation}
This leads to $\dot{V}_i \le -\rho(t)\sum_{j=1}^i c_j e_j^2 + e_i e_{i+1} + \sum_{j=1}^i e_j \Delta_j$.

\textbf{Step N (Final Control Law):} The composite Lyapunov function is $V_N = V_{N-1} + \frac{1}{2}e_N^2 + \frac{1}{2}\tilde{\theta}^T \Gamma^{-1} \tilde{\theta}$, where $\tilde{\theta} = \theta - \hat{\theta}$ and $\Gamma > 0$ is the adaptation gain matrix. The actual control input $u$ and the adaptive law are designed as:
\begin{subequations}\label{eq:control_laws}
\begin{equation}\label{eq:control_u}
\begin{split}
    u = \frac{1}{B_{N,1}} \Bigl( & -c_N \rho(t) e_N - e_{N-1} - (Az)_N - (Bu)'_N \\
    & + \dot{\alpha}_{N-1} - \hat{\theta}^T \Phi_N(z) \Bigr)
\end{split}
\end{equation}
\begin{equation}\label{eq:adapt_law}
    \dot{\hat{\theta}} = \Gamma \Phi_N(z) e_N
\end{equation}
\end{subequations}

where it is assumed $B_{N,1} \neq 0$, and $(Bu)'_N$ represents the part of $(Bu)_N$ not containing $B_{N,1}u$.

\section{Comprehensive Stability Analysis}

The composite Lyapunov function for the entire closed-loop system is:
\begin{equation}\label{eq:V_comp}
    V = \sum_{i=1}^N \frac{1}{2} e_i^2 + \frac{1}{2}\tilde{\theta}^T \Gamma^{-1} \tilde{\theta}
\end{equation}
Taking the derivative of $V$ and substituting the control and adaptive laws yields:
\begin{equation}\label{eq:V_dot_final}
    \dot{V} \le -\rho(t) \sum_{i=1}^N c_i e_i^2 + \sum_{i=1}^N e_i \Delta_i
\end{equation}
Using Young's inequality and Assumption 1, and after some algebraic manipulation (see Appendix A for the bound on $||z||$), we can show that:
\begin{equation}\label{eq:V_dot_bound}
    \dot{V} \le -2\rho(t) c_{\min} V_e - k_v V + D_0
\end{equation}
where $V_e = \frac{1}{2}||e||^2$, $c_{\min} = \min\{c_i\}$, $k_v > 0$ is a constant achievable by proper selection of design parameters, and $D_0$ is a positive constant that depends on the uncertainty bounds $\delta_0$ and $\delta_1$.

The inequality $\dot{V} \le -k_v V + D_0$ guarantees that $V(t)$ is Uniformly Ultimately Bounded (UUB). Furthermore, the term $-2\rho(t)c_{\min}V_e$ ensures that the tracking error $e(t)$ converges to a small residual set determined by $D_0$ within the prescribed time $T$, thus proving PPTS. This analysis explicitly links the ultimate control performance to the quality of the data-driven model via the uncertainty bounds.

\section{Simulation and Verification}
We validate our approach on the Van der Pol oscillator:
\begin{equation}\label{eq:vdp}
\begin{cases}
    \dot{x}_1 = x_2 \\
    \dot{x}_2 = \epsilon(1 - x_1^2)x_2 - x_1 + u + d(t)
\end{cases}
\end{equation}
where the damping coefficient $\epsilon=1$ is treated as the unknown parameter $\theta$. The objective is to stabilize the system to the origin $[0,0]^T$ with $d(t)=0.1\sin(\pi t)$. The system was excited with a pseudo-random binary sequence to collect data. A dictionary of 10 Radial Basis Functions (RBFs) was used for EDMDc. The prescribed time was set to $T=5$s.

\begin{figure}[t!] 
\centering
\includegraphics[width=\columnwidth]{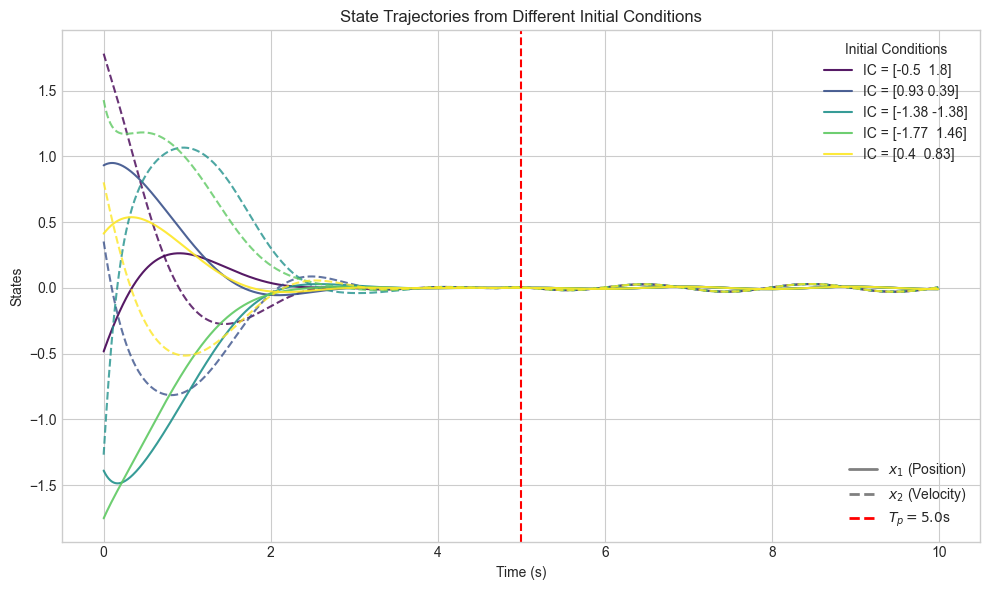} 
\caption{State trajectories of the Van der Pol oscillator under the proposed controller for five different initial conditions. All trajectories converge to a small neighborhood of the origin before the prescribed time $T=5$s (dashed line).}
\label{fig:vdp_states}
\end{figure}

The performance of the proposed controller is validated through simulations on the Van der Pol oscillator, with results from five different initial conditions depicted in Fig.~\ref{fig:vdp_states}. The figure clearly shows that both the position state ($x_1$, solid lines) and the velocity state ($x_2$, dashed lines) converge to a small neighborhood of the origin. Crucially, this convergence is achieved for all trajectories before the user-specified prescribed time $T=5$s, marked by the vertical dashed line. This result visually confirms the controller's effectiveness and its robustness to varying initial conditions, demonstrating its ability to enforce the prescribed-time stability objective on the nonlinear system.

\begin{figure}[t!]
\centering
\includegraphics[width=\columnwidth]{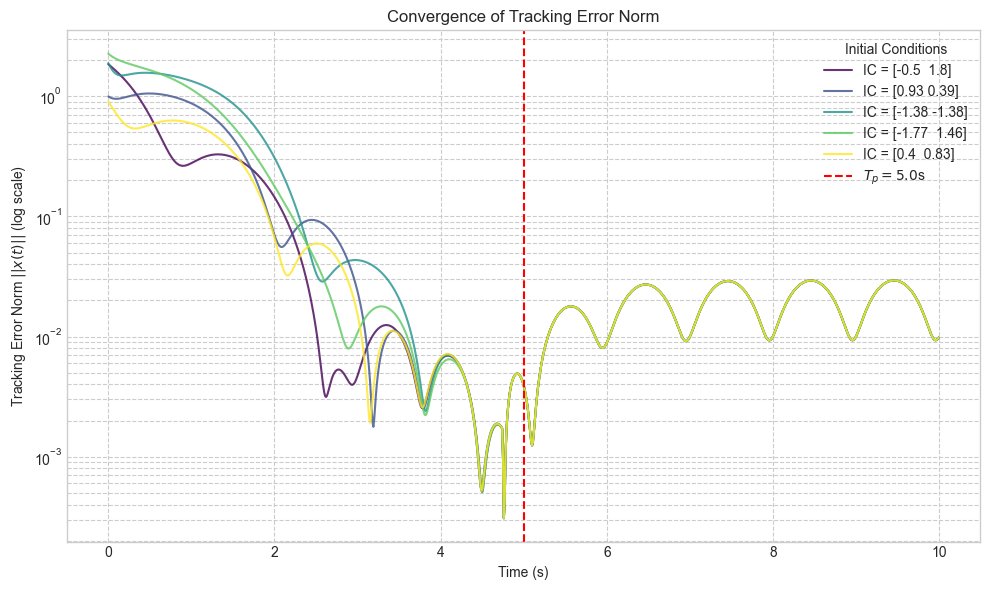} 
\caption{Convergence of the tracking error norm $||e(t)||$. The error enters and remains within a small residual set before $T=5$s.}
\label{fig:vdp_error}
\end{figure}

Figure~\ref{fig:vdp_error} provides a quantitative analysis of the convergence by plotting the Euclidean norm of the tracking error, $||e(t)||$, on a logarithmic scale. It can be observed that all error norms, regardless of their initial magnitude, decrease rapidly and enter a small residual set before the prescribed time $T=5$s. The logarithmic scale highlights the rapid rate of convergence, showing a reduction of several orders of magnitude. After the prescribed time, the error remains bounded within this set, which quantitatively verifies the practical prescribed-time stability (PPTS) property of the closed-loop system and its ability to counteract persistent disturbances.

\begin{figure}[t!]
\centering
\includegraphics[width=\columnwidth]{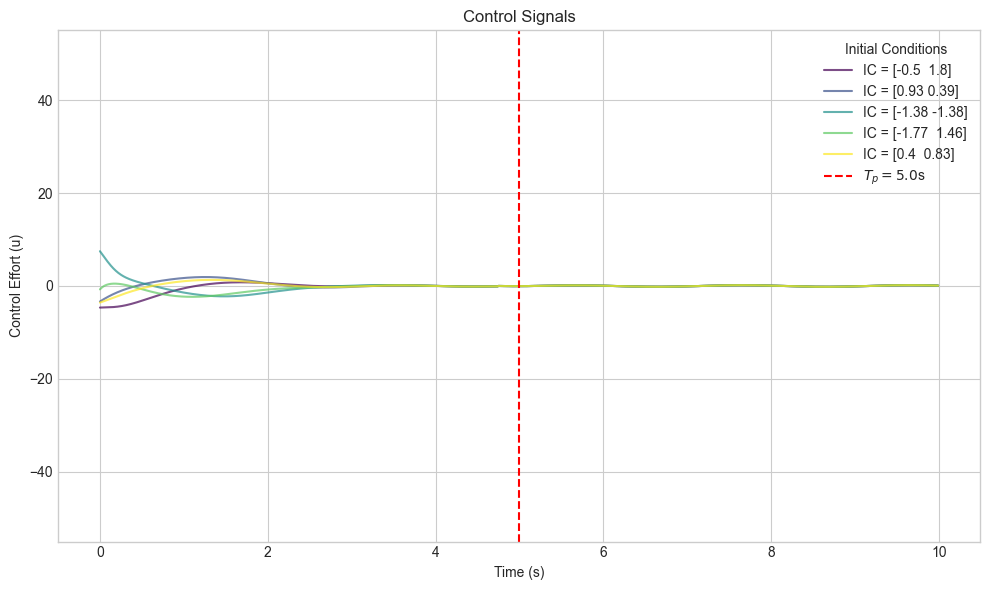} 
\caption{Control input signals, which remain bounded throughout the operation.}
\label{fig:vdp_input}
\end{figure}

The feasibility of the proposed control law is demonstrated in Fig.~\ref{fig:vdp_input}, which shows the control effort, $u(t)$, over time. The control signals are smooth and remain well within reasonable bounds throughout the simulation for all initial conditions. This is a critical result, as it confirms that the controller does not demand infeasible or instantaneous changes in actuation, making it suitable for practical implementation. After the system has stabilized, the control effort reduces to a small, oscillatory signal, which is precisely the action required to continuously reject the sinusoidal external disturbance, further illustrating the controller's robustness.

The simulation results in Figs. \ref{fig:vdp_states}-\ref{fig:vdp_input} clearly show that the system states are driven to a small neighborhood of the origin within the prescribed time $T=5$s, regardless of the initial conditions. The tracking error converges rapidly, and all signals remain bounded. These results are highly consistent with our theoretical analysis and demonstrate the effectiveness and robustness of the proposed Koopman-PTAB framework.

\section{Conclusion and Future Work}
This paper has presented a novel data-driven control framework that successfully integrates Koopman operator theory with prescribed-time adaptive backstepping. By designing the controller in the lifted linear space, we systematically addressed model uncertainty and circumvented the "explosion of complexity." Our rigorous Lyapunov analysis formally proves that the closed-loop system achieves practical prescribed-time stability, even with state-dependent uncertainties arising from data-driven modeling.

Future research directions include: (1) developing online learning schemes with adaptive basis functions to enhance adaptability to time-varying systems; (2) establishing more formal, probabilistic error bounds for the data-driven model for safety-critical applications; (3) extending the methodology to a broader class of systems, including MIMO systems; and (4) validating the algorithm on real-world physical systems.

\appendices
\section{Proof of the Bound on the Lifted State Norm}
\textbf{Lemma A.1:} Under the controller in Section III-B and Assumption 1, there exist positive constants $\gamma_1, \gamma_2, \gamma_3$ such that the squared norm of the lifted state, $||z||^2$, satisfies:
\begin{equation}\label{eq:z_bound_lemma}
    ||z||^2 \le \gamma_1 ||e||^2 + \gamma_2 ||\tilde{\theta}||^2 + \gamma_3
\end{equation}

\textit{Proof:} The proof proceeds by induction.
\textbf{Base case (i=1):} We have $z_1 = e_1 + z_{d,1}$. Since $z_d(t)$ is a bounded reference trajectory, $|z_{d,1}| \le Z_{d1, \max}$ for some constant $Z_{d1, \max}$. Thus, $z_1^2 = (e_1 + z_{d,1})^2 \le 2e_1^2 + 2z_{d,1}^2 \le 2e_1^2 + 2Z_{d1, \max}^2$. The lemma holds for $i=1$.

\textbf{Inductive step:} Assume the lemma holds for all $j \le i$. We have $z_{i+1} = e_{i+1} + \alpha_i$. Thus, $z_{i+1}^2 \le 2e_{i+1}^2 + 2\alpha_i^2$. The virtual control $\alpha_i$ is a function of $e_1, \dots, e_i$, $z_1, \dots, z_N$, $\hat{\theta}$, and bounded reference signals. Since $\hat{\theta} = \theta - \tilde{\theta}$ and $\theta$ is a bounded constant, and by the inductive hypothesis, the norm of $z_1, \dots, z_i$ is bounded by functions of $||e||_{[1, i-1]}^2$ and $||\tilde{\theta}||^2$, it follows that $\alpha_i^2$ can be bounded by a function of $||e||_{[1, i]}^2$ and $||\tilde{\theta}||^2$. Consequently, $z_{i+1}^2$ is bounded by a function of $||e||_{[1, i+1]}^2$ and $||\tilde{\theta}||^2$ plus a constant term.

By induction over $i=1, \dots, N$, we conclude that the lemma holds for the entire vector $z$. \hfill $\blacksquare$

\end{document}